\documentclass[preprint,amsmath,amssymb,aps,showkeys,showpacs,pra,]{revtex4-1}

\usepackage[english]{babel}
\usepackage{graphicx}  
\usepackage{amsmath}
\usepackage{dcolumn}
\usepackage{amssymb}
\usepackage{bm}
\usepackage[latin1]{inputenc}
\usepackage{graphicx,color}
\usepackage[pdftex]{hyperref}

\begin{document}
\title{Two-Dimensional  Quantum ring in a Graphene Layer in the presence of a Aharonov-Bohm flux}
\author{José Amaro Neto, M. J. Bueno and Claudio Furtado}
\email{furtado@fisica.ufpb.br}
\affiliation{Departamento de F\'isica, Universidade Federal da Para\'iba, Caixa Postal 5008, 58051-970, Jo\~ao Pessoa, PB, Brazil.} 

\begin{abstract}
In this paper  we study the relativistic  quantum dynamics of a massless fermion confined in a quantum ring.  We use a model of confining potential and introduce  the interaction via   Dirac oscillator coupling,  which provides ring confinement for massless Dirac fermions.  The energy levels and   corresponding eigenfuctions for this model  in graphene layer in the presence of Aharonov-Bohm flux in the centre of the ring and the expression for persistent  current in this model are derived. We also investigate the  model for quantum ring in graphene layer in the presence of  a disclination and  a magnetic flux.  The  energy spectrum and wave function are obtained exactly for this case.  We see that the persistent current depends  on parameters  characterizing the topological defect.
\end{abstract}


\maketitle

\section{Introduction}
With the experimental obtaining  of graphene \cite{novo},  several interesting physical phenomena were observed in this material, such as:  anomalous Hall effect \cite{kim}, Klein paradox ~\cite{Novoselov3,Young,stan}, spin  qubits~\cite{bula}, Moir\'e potential~\cite{Sub}, etc. Experiments involving the study of quantum dots in graphene in the presence of magnetic field were performed \cite{pono, schn1},  in which the energy spectrum in this system  was observed. Theoretical models of quasiparticles confined  in quantum dots\cite{silv, schn2} and rings\cite{peters1,peters2,peters3} in graphene have been recently proposed.  In a mesoscopic electronic system, it is well known that when  one  varies the magnetic flux  passing through the   centre of device  of a ring topology, a persistent current  arises due to the Aharonov-Bohm effect~\cite{Bleszyski}. This quantum coherence plays a central role in mesoscopic physics.  From the theoretical point  of view the persistent current has been investigated recently theoretically in Ref. ~\cite{TanInkson, Guinea, Ma,eplfur,linco1,linco2} and experimentally in ~\cite{Eckern}.

The Dirac oscillator \cite{Moshinsky} is  introduced as the relativistic version of the harmonic oscillator due  to the fact that  this coupling produces in the non-relativistic limit a harmonic oscillator with a strong spin-orbit coupling.  The Dirac oscillator coupling is introduced via the following  replacement in Dirac equation: $ {p} \rightarrow {p} - iM\omega\widehat{\beta}\vec{x}$, where $M$ is  a mass of  the particle, and $\omega$ is the frequency of the oscillator. The Dirac oscillator has been investigated in several areas of physics: in the study of the hidden supersymmetry  by Benitez {\it et al} \cite{benitez};  within  the  analogy  with the Jaynes-Cummings model  which was  demonstrated by Rozmej and  Arvieu \cite{rozme}; in the presence of topological defects \cite{josevi}. Additionally, Bermudez {\it et al.} \cite{bermu}  have studied the Ramsey  interferometric effect in non-relativistic limit of the Dirac oscillator. Recently,  applications of Dirac oscillator  have been carried out in graphene by Quimbay and Strange \cite{strange}, in  studies of quantum dots by Belouad {\it et al.} \cite{jellal} performed with use of the confining models proposed in Ref. \cite{jellal1,jellal2}. Thermal properties of the Dirac oscillator  have been investigated by Boumali and  Hassanabadi \cite{boumeplp}.  Recently, the Dirac oscillator was observed experimentally  in  microwave physics \cite{franco}.   The  quantum dot in graphene   in the presence {\bf of a} topological defect was studied by Bueno {\it et al.} \cite{Bueno}, and spectrum of energy and the current  persistent are obtained. In Ref. \cite{Bueno} the confinement potential was coupled in Dirac equation in the way  similar to Dirac oscillator \cite{Moshinsky}. In Ref.\cite{Bueno1}, the Landau levels in a graphene layer with the presence of disclination was investigated.

In this paper  we investigate the quantum ring in  a graphene layer. We use the model of confining potential proposed in Ref. \cite{Bakke}, this model is  a relativistic version of the  Tan-Inkson model  for the confining potential in  two-dimensional space \cite{TanInkson}. We investigate, in low energy limit, the system described by a massless  Dirac equation, where  a continuous description  near Fermi $K$-points is employed.   We use the Dirac oscillator type coupling to confine harmonically the  quasiparticles  in quantum ring pierced by Aharonov-Bohm flux in a graphene layer.   We  also  study the influence of a disclination in two-dimensional  quantum ring of nanometric size in the presence of Aharonov-Bohm quantum flux.  From the  Landau levels in graphene layer  it is possible to obtain the  magnitude of  magnetic length $\ell_{b}=\sqrt{\hbar/eB} \approx 50 nm$. This fact  demonstrates the relevance  of studying the physical  influence of {\bf a} disclination in this quantum ring,  keeping in mind that the average size of one topological defect of this kind is of the order of interatomic distances  for the carbon atoms in this structure.  In both cases, we obtain  the eigenvalues and eigenfunctions of energy and  persistent current.  In the case of dynamics in the presence of defects, we  demonstrate the dependence of these physical quantities  on the parameter  characterizing the disclination, thus demonstrating the influence of the defect in the  dynamics in a quantum ring.

This paper is organized as follows: in section II, we present  the  confining model  and  the corresponding coupling in Dirac equation. In Section   III, we analyze   the quantum dynamics of  massless Dirac fermions in a quantum ring in graphene layer, and  obtain the  exact energy spectrum  and eigenfunctions for  the  model  where a thin Aharonov-Bohm flux confined to the center of the ring is introduced,  and the  persistent current is  obtained in this  case. In section IV, we obtain the eigenvalues and eingefunctions for quantum dynamics of massless Dirac fermions in a quantum ring in the presence of  a disclination,  and study the influence of topological defect is investigated and the persistent current, and finally, in section V we present the concluding remarks.

\section{The Ring Confinement  Model}\label{sec2}
In this section we present the potential used to confine the quasiparticle in a  ring-like topology in graphene layer. Recently in Ref. \cite{Bakke}  the authors proposed an extension to relativistic case of Tan-Inkson model \cite{TanInkson}, which has been constructed to model quantum rings  within non-relativistic dynamics of  electrons and holes. In this relativistic  model of confinement the authors \cite{Bakke}  also used two control parameters to obtain a harmonic confinement in a two-dimensional ring, and the quantum point limits are obtained when we make one of the parameters to be zero, $a_1=0$. In this relativistic model the confining potential is introduced via coupling  of  the momentum  of a quasiparticle  in a manner similar to Dirac oscillator \cite{Moshinsky}.    In  the non-relativistic limit, the Tan-Inkson  potential is obtained. The coupling used in the massless Dirac equation to obtain the harmonic confining potential in ring topology in two-dimensional relativistic system  is given by

\begin{equation}
\mathbf{p} \rightarrow \mathbf{p}  + i \left[ \frac{\sqrt{2a_1}}{\rho} + \sqrt{2a_2}\rho  \right] {\gamma}^{0}\widehat{e}_{\rho} \label{eq1},
\end{equation}
where $a_1$ and $a_2$ are  the parameters of the potential. Note that  at $a_{1}\longrightarrow 0$ we obtain a quantum dot harmonic confining potential.
For the case $a_{2} \longrightarrow 0$ we obtain the  relativistic antidot. Now we use this new coupling in {\bf a} massless Dirac equation to describe a quantum ring structure in graphene  in the low-energy limit.

\section{The Quantum Ring in a Graphene Layer}
The graphene physics can be treated as  a laboratory for  studies in relativistic quantum mechanics in (2+1)-dimensions. In the low energy limit near the Fermi  points $\bf{ K}_{+}$ and  $\bf{ K}_{-}$, the dynamics of  quasiparticle is described   by  a relativistic massless fermionic theory.  In this theory, the massless Dirac spectrum arises because of the lattice structure of graphene. The hexagonal structure of graphene lattice has a basis  composed by two  sets of carbon atoms, this symmetry  is responsible for the origin of the degree of freedom of isospin.  In this  energy scale the graphene layer is described by the  Dirac equation. Now   we introduce the Dirac equation in cylindrical  coordinates.   We start from the  Dirac equation in general coordinate system \cite{Fock}, given by
\begin{equation}
i{\gamma}^{\mu}D_{\mu}\Psi + \frac{i}{2}\sum_{k=1}^{3}{\gamma}^{k}\left[D_k \ln\left(\frac{h_1 h_2 h_3}{h_k}\right) \right]\Psi = 0,
\label{eq2}
\end{equation}
 where  we defined $v_{f}=\hbar=1$, where $v_{f}$ is the Fermi velocity,  and $k =1,2,3$. In this equation, $D_{\mu} = \frac{1}{h_{\mu}}  \frac{\partial}{\partial {x}^{\mu}}$ corresponds to  the derivative in  the general coordinate system and the  parameters  $h_k$ are the scale factors. Assuming the cylindrical symmetry, one can write  metric in Minkowski space-time  as  
$ds² = -dt² + d{\rho}² +{\rho}²d{\varphi}² $, from which we obtain the scale factors: $h_0= 1,  h_1 = 1,  h_2 = \rho $; and the coordinates: $ x^{0} = t, x¹ = \rho , x²=\varphi$ .  The matrices ${\gamma}^{\mu}$ are the Dirac matrices defined in Minkowski space-time like:
\begin{eqnarray}
{\gamma}^{0} =\widehat{\beta}= \left( \begin{array}{ccc}
I & 0&  \\
0 & -I\end{array}\right), \quad {\gamma}^{i} =\widehat{\beta}\widehat{\alpha}^{i} =\left( \begin{array}{ccc}
0 & {\sigma}^{i}&  \\
-{\sigma}^{i} & 0\end{array}\right), \label{eq3}
\end{eqnarray}
 with $I$ and ${\sigma}^{i}$  are $2 \times 2$ matrices, being the identity and Pauli matrices respectively. The matrices $\widehat{\alpha}^{i}(i = 1,2,3)$ and $\widehat{\beta}$ satisfy the set of properties
\begin{align}
 & \widehat{\alpha}^{i}\widehat{\alpha}^{j} +
\widehat{\alpha}^{j}\widehat{\alpha}^{i} = 2{\delta}_{ij}I 
\nonumber \\ 
& \widehat{\alpha}^{i}\widehat{\beta} =
- \widehat{\beta} \widehat{\alpha}^{i}
 \nonumber \\
&{\widehat{\alpha}^{i}}² = {\widehat{
\beta} }² =I \label{eq4}
\end{align}
In graphene, in regions  near the Fermi points, the electrons behave like  massless fermions, so that (\ref{eq2}) is rewritten as
\begin{equation}
i{\gamma}^{0} \frac{\partial \Psi}{\partial t} +i{\gamma}^{1} \frac{\partial \Psi}{\partial \rho}+i\frac{{\gamma}^{2}}{\rho} \frac{\partial \Psi}{\partial \varphi}+\frac{i{\gamma}^{1}}{2 \rho}\Psi= 0.
\label{eq5}
\end{equation}
Using this equation, taking into account the properties of the matrices (\ref{eq3}) and \ref{eq4}, the  Dirac equation  can be presented as
\begin{equation}
i\frac{\partial \Psi}{\partial t} = \left[ -i{\alpha}¹\left(\frac{\partial}{\partial \rho} + \frac{1}{2 \rho}\right)
-\frac{i{\alpha}²}{\rho}\left(\frac{\partial}{\partial \varphi }\right)  \right]\Psi, \label{eq6}
\end{equation}
where we write the  Dirac-type Hamiltonian only as $\widehat{H} = \widehat{\alpha} \mathbf{p}$. 

Now let us consider a Aharonov-Bohm flux  passing  through  the centre of the ring,   and a field described by the vector potential $\mathbf{A} = \frac{\Phi}{2\pi \rho}\widehat{e}_{\varphi}  $ . The  Aharonov-Bohm flux (magnetic flux) $\Phi$ \cite{Aharonov} goes along the $z$-direction. We use the minimal  coupling in the  momentum $\Pi = \mathbf{p} - |q|\mathbf{A}$  thus changing $\mathbf{p}$ for $\Pi $ so that $ \widehat{H} =\widehat{ \alpha} (\mathbf{p} - |q| \mathbf{A})$.
Thus the Hamiltonian of graphene in the presence of the magnetic field with the coupling (\ref{eq1}) is written as
\begin{equation}
\widehat{H}_B = \left[ -i{\alpha}¹\left(\frac{\partial}{\partial \rho} + \frac{1}{2 \rho}-\widehat{\beta} \frac{\sqrt{2a_1}}{\rho} - \widehat{\beta} \sqrt{2a_2}\rho \right)
-\frac{i{\alpha}²}{\rho}\left(\frac{\partial}{\partial \varphi } - i\frac{\Phi}{{\Phi}_0}\right)  \right] \label{eq7} 
\end{equation}
with ${\Phi}_0 = \frac{2\pi}{|q|}$ is the quantum Aharonov-Bohm magnetic flux. The solution of the  Dirac equation is defined by \emph{ansatz}:
\begin{equation}
\Psi = e^{-iE t}\left(\begin{array}{c}{\psi} \\ {\phi} \end{array}\right)\label{eq8},
\end{equation}
where the spinors ${\psi} = {\psi} (\rho,\varphi)$ and ${\phi} = {\phi}(\rho,\varphi)$ represent each sublattice of  the graphene monolayer. Applying the Hamiltonian (\ref{eq7}) in the \emph{ansatz} (\ref{eq8}), we have two coupled equations for ${\psi}$ and ${\phi}$:
\begin{eqnarray}
E{\psi} &= & -i{\sigma}¹\left[ \frac{\partial}{\partial \rho} + \frac{1}{2 \rho}+ \frac{\sqrt{2a_1}}{\rho} + \sqrt{2a_2}\rho\right]{\phi} -i\frac{{\sigma}²}{\rho}\left[\frac{\partial}{\partial \varphi } - i\frac{\Phi}{{\Phi}_0}  \right]{\phi} 
\label{eq9}\end{eqnarray} 
and
\begin{eqnarray}
E {\phi} &= & -i{\sigma}¹\left[ \frac{\partial}{\partial \rho} + \frac{1}{2 \rho}- \frac{\sqrt{2a_1}}{\rho} - \sqrt{2a_2}\rho\right]\psi -i\frac{{\sigma}²}{\rho}\left[\frac{\partial}{\partial \varphi } - i\frac{\Phi}{{\Phi}_0}\right]\psi,
\label{eq10}\end{eqnarray}
thus, eliminating $\phi$,   and substituting (\ref{eq9}) in (\ref{eq10}), we obtain the second-order differential equation:
\begin{eqnarray}
E² \psi &=&-  \left( 1+ \frac{1}{{\rho}²}\right)\frac{{\partial}^2\psi}{\partial {\rho}²}-\frac{1}{\rho}\frac{\partial\psi}{\partial \rho} 
+\left(\frac{1}{4}+ 2a_1 - \sqrt{2a_1}+\frac{\Phi²}{\Phi_0²}\right) \frac{\psi}{{\rho}²} + 2a_2 {\rho}²\psi + \nonumber\\ &+& \left(\sqrt{2} + 4\sqrt{a_1}\right) \sqrt{a_2}\psi \nonumber\\ 
&+& \frac{2}{{\rho}²}\frac{\Phi}{{\Phi}_0}\frac{i\partial \psi}{\partial \varphi} 
+{\sigma}³\left[\left( \frac{1}{{\rho}²}- \left(\frac{\sqrt{a_1}}{\rho}+ \sqrt{a_2}\right)2\sqrt{2}\right)\left(\frac{\Phi}{{\Phi}_0}+\frac{i\partial }{\partial \varphi}\right)\right]\psi.   
\label{eq11}\end{eqnarray}
In (\ref{eq11})  it is possible to verify  that $\psi$ is the eigenfunction  of ${\sigma}^3$ whose eigenvalues are $s=\pm1$, where we write ${\sigma}^3{\psi}_s =\pm{\psi}_s=s{\psi}_s $. We can also see that the operators $\widehat{J}_z=-i\frac{\partial}{\partial\varphi}$ and $\widehat{p}_z =-i\frac{\partial}{\partial z}$  are  compatible with the conserved observables, i.e.,  they both commute with the Hamiltonian of the right side of equation, so we  can take solution of (\ref{eq11}) in  the form:
\begin{equation}
\psi = e^{ij\varphi}\left(\begin{array}{c}R_+ (\rho) \\ R_{-} (\rho)  \end{array}\right)\label{eq12},
\end{equation} 
where $j= l + \frac{1}{2} $, with $l= 0,\pm 1,\pm 2,..$. Substituting (\ref{eq12}) in (\ref{eq11}),  using ${\sigma}³{\psi}_s = \pm{\psi}_s = s{\psi}$, and considering   the property  of Pauli  matrices  that  ${\sigma}^{i}{\sigma}^{i} = I$ and the notation $R_s(\rho) = (R_{+}(\rho),R_{-}(\rho)) $, we obtain the  following  radial equation:
\begin{equation}
\left[ \frac{d²}{d{\rho}²} + \frac{1}{\rho}\frac{d}{d\rho} - \frac{{{\vartheta}_s}²}{{\rho}²} - 2a_2{{\rho}²}+{\varepsilon}_s \right]R_s (\rho) = 0\label{eq13},\end{equation} 
where we define the following  parameters:
\begin{eqnarray} {\vartheta}_s &=& {\varsigma}_s + s\sqrt{2a_1}  , \\ {\varsigma}_s &=& l + \frac{1}{2}(1-s)- \frac{\Phi}{{\Phi}_0}, \nonumber\\
{\varepsilon}_s &=& {E}² - 2s\sqrt{2a_2}{\varsigma}_{s} - 2\sqrt{2a_2} - 4\sqrt{a_1 a_2} . \label{eq14} 
\end{eqnarray}
To solve equation (\ref{eq13}), we use the  change of  the variable $\varrho  = \sqrt{2a_2}{\rho}²$,  and  obtain
\begin{equation}
\left[ \varrho \frac{d²}{d{\varrho}²} + \frac{d}{d\varrho} - \frac{{{\vartheta}_s}²}{4 \varrho} - \frac{{\varrho}}{4} +\frac{{\varepsilon}_s}{4\sqrt{2a_2}} \right]R_s (\varrho) = 0 \ .\label{eq15}\end{equation} 
The solution of  the equation (\ref{eq15}) must be regular at the origin and finite everywhere, i.e., $R_s \rightarrow 0$  where $\rho \rightarrow \infty$. The function  satisfying these conditions has the form
\begin{equation}
R_s(\varrho)= e^{\frac{-\varrho}{2}}{\varrho}^{\frac{|{\vartheta}_s|}{2}}F_s (\varrho),\label{eq16}
\end{equation}
and substituting (\ref{eq16}) in( \ref{eq15}), we  arrive at  following equation
\begin{equation}
 \varrho \frac{d²F_s}{d{\varrho}²} +[ |\vartheta | + 1 - \varrho] \frac{dF_s}{d \varrho}  + \left[\frac{{\varepsilon}_s}{4\sqrt{2a_2}} -\frac{|{\vartheta}_s|}{2} - \frac{1}{2} \right]F_s (\varrho) = 0 , \label{eq17}\end{equation}
which corresponds to confluent hypergeometric  equation \cite{Abramo},  whose solution  is given by
\begin{equation} 
F_s ={\phantom{1}_1}F_1(a,b;z)=
 {\phantom{1}_1}F_1\left(\frac{|{\vartheta}_s|}{2} + \frac{1}{2} -\frac{{\varepsilon}_s}{4\sqrt{2a_2}},|{\vartheta}_s|+1,\varrho \right). 
 \end{equation}
Imposing the condition  that the solution  for  hypergeometric series becomes a polynomial of degree $n$, i.e., $ \frac{|{\vartheta}_s|}{2} + \frac{1}{2} -\frac{{\varepsilon}_s}{4\sqrt{2a_2}} = -n$, 
and using the definitions of  parameters (\ref{eq14}), we obtain the following eigenvalues 
\begin{eqnarray}
{E}_{n,l}  & = &\pm  \left\{4\sqrt{2a_2}\left[n + \frac{|l + \frac{1}{2}(1-s)- \frac{\Phi}{{\Phi}_0} +s\sqrt{2a_1}|}{2} + s\frac{(l + \frac{1}{2}(1-s) -\frac{\Phi}{{\Phi}_0})}{2} +1 \right]+\right.\nonumber\\& &\left. {}+ 4\sqrt{a_1 a_2}\right\}^{1/2},
\label{eq19}
\end{eqnarray}
 which is the spectrum of energy for the quasiparticle($E>0$ is associated electrons and $E <0$ with holes) confined in the two-dimensional quantum ring on a monolayer of graphene. Note   that  the energy spectrum \ref{eq19}  depends  on the control parameters $a_1$ and $ a_2 $ and the quantum numbers $ n $ and $ l $. The energy spectrum depends on the  parameter characterizing  the quantum ring in the graphene layer and on the  Aharonov-Bohm flux. In the limit $ a_1 \longrightarrow 0 $  in Eq. (\ref{eq19}), we obtain the spectrum of quasiparticle in quantum dot in graphene layer.

The solutions  with positive energy $E>0$ describe the dynamics of electrons in the conduction band, while  $E <0$ describes the dynamics of holes. Now we obtain the spinors corresponding to positive energy in    (\ref{eq19}).  First  of all, we should note that, by  using the radial wave functions (\ref{eq16}), we can write (\ref{eq12}) in the following form
\begin{equation}
 {\psi}_s =e^{ij\varphi} R_{s}(\varrho)= e^{i(l+1/2)\varphi} e^{-\sqrt{\frac{a_2}{2}}{\rho}^{2}} (2a_2 )^{\frac{|{\vartheta}_s|}{4}} {\rho}^{|{\vartheta}_s|} \times {\phantom{1}_1}F_1 \left(-n,|{\vartheta}_s|+1, \sqrt{2a_2}{\rho}²\right)\label{eq21}. 
 \end{equation}
Substituting   (\ref{eq21}) into (\ref{eq9}), we obtain the solution for spinors of $\phi$ given by
\begin{eqnarray}
{\phi} =& \frac{i}{E}N_s\Bigg\lbrace  \left[2\sqrt{2a_2}\rho - \frac{|{\vartheta}_s|}{\rho} -\frac{1}{2\rho} + \frac{{\sqrt{2a_1}}}{\rho}   \right]{\sigma}¹{\phantom{1}_1}F_1 \left(-n,|{\vartheta}_s|+1, \sqrt{2a_2}{\rho}²\right)\Bigg \rbrace \nonumber \\ & +\frac{i}{E}N_s\Bigg\lbrace \frac{2 n \sqrt{2a_2}\rho}{|{\vartheta}_s|+1}  {\sigma}¹{\phantom{1}_1}F_1 \left(-n + 1 ,|{\vartheta}_s|+2, \sqrt{2a_2}{\rho}²\right)\Bigg \rbrace \nonumber \\&    + \frac{1}{E}N_s\Bigg\lbrace \left[   \frac{1}{\rho}\left(l + \frac{1}{2} - \frac{\Phi}{{\Phi}_0} \right) \right]{\sigma}²      {\phantom{1}_1}F_1 \left(-n,|{\vartheta}_s|+1, \sqrt{2a_2}{\rho}²\right) \Bigg \rbrace  \label{eq22},
\end{eqnarray} 
where $N_s$ is the  normalization constant. In this way, the  bispinors can be represented as
\begin{eqnarray}
\psi(\rho)= \left(\begin{array}{c} {\psi}_{+}(\rho)\\ {\psi}_{-}(\rho) \end{array}\right)\label{eq23}.
\end{eqnarray}
From equations (\ref{eq8}), (\ref{eq22}) and (\ref{eq23}),   it is  possible  to write the positive energy solutions of the Dirac equation corresponding to the parallel and antiparallel components toward $z$  axis:
\begin{eqnarray}
{\Psi}_+ =& f_+ {\phantom{1}_1}F_1 \left(-n,|{\vartheta}_+|+1, \sqrt{2a_2}{\rho}²\right) \times 
\left( \begin{array}{c} 1 \\ 0 \\ 0 \\ \frac{i}{E}\left[2\sqrt{2a_2} \rho + \frac{\left( {\varsigma}_+ - |{\vartheta}_+| + \sqrt{2a_1} \right) }{\rho}\right]
\end{array} \right)\nonumber \\&+ \frac{i f_+}{E}\left(\frac{2n\sqrt{2a_2}\rho}{|{\vartheta}_+|+1} \right){\phantom{1}_1}F_1 \left(-n + 1 ,|{\vartheta}_{+}|+2, \sqrt{2a_2}{\rho}²\right)\left( \begin{array}{c} 0 \\ 0 \\ 0 \\ 1\end{array} \right)\label{eq24}
\end {eqnarray} 
for $s= +1$ e ${\psi}_{-}(\rho) = 0 $, by analogy,
\begin{eqnarray}
{\Psi}_{-} = & f_{-} {\phantom{1}_1}F_1 \left(-n,|{\vartheta}_{-}|+1, \sqrt{2a_2}{\rho}²\right) \times 
\left( \begin{array}{c} 0 \\ 1 \\ \frac{i}{E}\left[2\sqrt{2a_2} \rho - \frac{ \left( {\varsigma}_{-} + |{\vartheta}_{-}| - \sqrt{2a_1} \right)}{\rho}\right] \\ 0
\end{array} \right)\nonumber \\& + \frac{i f_{-}}{E}\left(\frac{2n\sqrt{2a_2}\rho}{|{\vartheta}_{-}|+1} \right){\phantom{1}_1}F_1 \left(-n + 1 ,|{\vartheta}_{-}|+2, \sqrt{2a_2}{\rho}²\right)\left( \begin{array}{c} 0 \\ 0 \\1 \\ 0\end{array} \right) \label{eq25}, \end{eqnarray} 
for $s= -1$ e ${\psi}_{+}(\rho) = 0 $,
where the factor $f_s$ in equations (\ref{eq24}) and (\ref{eq25}), is written  as
\begin{eqnarray}
 {f}_s =\left( \frac{\sqrt{8a_2}\Gamma(|{\vartheta}_s|+n+1)}{\Gamma(n+1)\left[\Gamma(|{\vartheta}_s|+1)\right]²} \right)^{1/2}  \times e^{-iEt}e^{i(l+1/2)\varphi} e^{-\sqrt{\frac{a_2}{2}}{\rho}^{2}} (2a_2 )^{\frac{|{\vartheta}_s|}{4}} {\rho}^{|{\vartheta}_s|},
\label{eq26} \end{eqnarray} 
with $\Gamma(n)$ is   the Euler gamma function.  This solution correspond  to the $E>0$  case for massless Dirac equation for a quasiparticle confined in quantum ring in  a graphene layer.
\subsection{The Persistent Current for Quantum Ring in Graphene Layer}
 In this subsection we use the Byers-Yang relation~\cite{Byers} and the expression of  eigenvalues (\ref{eq19}) to calculate the persistent current for the quantum ring in graphene. With help of  this relation we can obtain the persistent  current by  differentiating the  expression of energy  with  respect to the magnetic flux   
\begin{eqnarray}\label{current}
I = -\sum_{n,l} \frac{\partial E_{n,l}}{\partial \Phi}.
\end{eqnarray}
Substituting equation (\ref{eq19}) into the equation (\ref{current}),  we can calculate the persistent currents in the system:
\begin{eqnarray}
I = \frac{|q|}{2 \pi}\sum_{n,l} \sqrt{2a_2} \left(  \frac{\pm{\vartheta}_s}{|{\vartheta}_s|} +1\right) \times \Bigg \{  4\sqrt{2a_2}\left[ n + \frac{|{\vartheta}_s|}{2} + s\frac{{\varsigma}_s}{2} +1  \right] + 4\sqrt{a_1 a_2} \Bigg \}^{-1/2},
\label{eq20}
\end{eqnarray}  
the  persistent current (\ref{eq20}) depends  on the control parameters $a_1$ and $a_2$ and the quantum numbers $n$ and $l$. The current is a periodic  function  of the Aharonov-Bohm flux $\Phi_{AB}$. Note that we have considered the persistent current for electrons $E>0$. For the case of holes we use similar calculation for the case $E <0$.

\section{Quantum Ring in Graphene Layer  with Disclination}

In this section we  study  the quantum ring in  a graphene layer in the presence of  a topological defect,  in this case, a disclination. We use the geometric theory of defects \cite{kat} to describe in the continuum limit  the graphene layer with  a topological defect.  In a real  crystal lattice, some topological  defects  can  occur. The study of such defects has shown a number of technological applications \cite{Lahiri},  as  well as better understanding  of  the electronic transport properties, phase  transitions and diffusion in condensed matter systems \cite{Hyde}. Among  all this, we highlight the topological  defects of a structural nature,  which have  been studied both in gravitation  and in  condensed matter. In both  cases, topological defects are associated  with symmetry properties  of the system.  Dislocations are associated to translational  symmetry of the lattice, and disclinations are associated with  rotational  symmetry.  From the geometric viewpoint, dislocations  are related with torsion of elastic medium and disclinations  are associated with curvature of this medium.  This topological defect can be obtained by Volterra \cite{Volterra} process. This process can be visualized by  ``cut and glue" procedure to obtain the topological defect.  The studies  of curved structures of carbon using massless Dirac  equation  intended to investigate electronic properties of these structures, have been  carried out by several authors. In particular,  Gonz\'alez, Guinea and Vozmediano~\cite{voz1,voz2} investigated a model  to describe  the fullerene molecules, in these studies the fullerene molecule is described by a spherical model in the presence on a non-Abelian gauge field produced by a magnetic monopole at the centre of the sphere. Recently, the electronic structure of a graphitic nanoparticle was  investigated by Osipov, Kochetov and Pudlak~\cite{osi1,osi2,osi3} using a field theory model. Osipov and collaborators investigated  disclinations in the   conical, spherical and hyperbolic geometries. The local density of states was obtained using the Dirac equation in these geometries.  It is  well known that the spinor  describing the quasiparticles near Fermi points  acquires  a holonomy  being transported around a disclination. This effect can be viewed as a geometrical Aharonov-Bohm effect ~\cite{crespi1,crespi2}. This result was generalized to a system with $n$ disclinations,  whose resulting configuration is described by an effective defect~\cite{alex2}.  The three-dimensional metric  describing a layer with topological defect in geometric theory\cite{kat} is given by
\begin{equation}
ds² = dt² - d{\rho}² - {\alpha}²{\rho}²d{\varphi}², 
\label{eq28}\end{equation} 
where  $\alpha$ is the intensity of  the disclination,  it can be written in terms of the angular sector $\theta$, which we removed or inserted in the graphene layer to form the defect,   as  $ \alpha =1 \pm \theta / 2 \pi$.  This cut and glue process  obeys the symmetry of the honeycomb lattice and the sector $\theta$  is a multiple of $\pi /3$,  so, $\theta=\pm N \pi /3$ where  $N$  is  an  integer   with $0<N< 6$. The   ``cut and glue" procedure to visualize the topological defect is named  the Volterra process,   when the  parameter $\alpha$ within the range $ 0 <\alpha<1$ characterizes   a positive disclination, where  within the  Volterra process we remove an angular sector $- N \pi /3$ of graphene layer.  The  $\alpha $ in the range $1<\alpha<\infty$ characterize negative disclinations, i.e.,  the defects  formed through insertion of an  angular sector $+ N \pi /3$   within the Volterra process. In  a graphene layer, disclinations described by the line element~(\ref{eq28})  correspond to removed sectors or inserted sectors. We can use gauge fields to describe the influence  of a disclination in the electronic properties of graphene.  First,  for the gauge field,  its  quantum fluxes measure the angular deficit of the cone when a spinor is  transported  in a parallel manner around the apex in a closed path proving us the variation of the local reference frame along the path.  The matrix of parallel transport or holonomy transformation ~\cite{alex2,crespi1,crespi2} are obtained of   geometry  (\ref{eq28})  and responsible by the variation of the local reference frame in the graphene layer and  produces a flux that acts in the sublattices  $A/B$ of graphene layer with defect and are expressed by following  integral 
\begin{equation}
\label{paralel2}
\oint \Gamma_{\mu}dx^{\mu}=\pi (\alpha-1)\sigma^{z}.
\end{equation}
We have another contribution for the quantum holonomy  in the graphene with topological defects  and named of the spin flux. This spin flux is responsible  for mixing of the Fermi points $K_{\pm}$\cite{pachos}, and  its expression is given by   
\begin{equation}
\label{kspin2}
\oint a_{\mu}dx^{\mu}=-3\pi (\alpha-1)\tau^{y}.
\end{equation}
where  $\tau^{i}$ are the Pauli matrices. Notice that the holonomy (\ref{kspin2}) acts in the $K$-space in contrast with the flux (\ref{paralel2}) which acts in the sublattices $A/B$. In this way, the $\tau^{i}$ Pauli matrices  act in the  $K$-space and the Pauli matrices $\sigma^{i}$  act in the real space. Note that the expressions (\ref{paralel2}) and (\ref{kspin2}) are functions of the parameter $\alpha$ characterizing the presence of a disclination.

In this way, the Dirac equation for  massless quasiparticles can be rewritten   in the presence of non-Abelian gauge field  introduced by  the disclination and described in  Eq. (\ref{kspin2}). This equation  is given by
\begin{equation}
\label{eq35}
\left[i\gamma^{\mu}\frac{\partial}{\partial x^{\mu}}-i\gamma^{\mu} \Gamma_{\mu}-\gamma^{\mu}\frac{a_{\mu}}{\rho}\right]\psi=0,
\end{equation}
where the $\gamma^{\mu}$  are the  Dirac matrices  defined in a curved space.  The  $\Gamma_{\mu}$ term  is   a spinor  connection,   being present due to the curved nature of the geometry of the disclinated layer of graphene in  a continuous limit. Moreover, we  note in (\ref{eq35}) the presence of a new term $a_{\mu}$ which represents a non-Abelian gauge field related to the $K$ spin flux. The non-zero non-Abelian gauge   field $a_{\mu}$  found  in (\ref{eq35})  is responsible for the mixing of  the Fermi-points  $K_{\pm}$.  Basing on the integral (\ref{kspin2}),  one can express the non-zero component  for the  non-Abelian gauge field $a_{\mu}$ in the graphene layer with disclination   as
\begin{equation}
a_{\phi}=\pm \frac{3}{2}\left(\alpha -1 \right),\label{eq37}
\end{equation}
where  the sign $\pm$  sign is related with  the  $K_{\pm}$ spin fluxes.
In a curved space  the $\gamma^{\mu}$ matrices can be expressed in function of triads  fields $e_{a}^{\mu}\left(x\right)$. In this curved background the Dirac matrices  must  be defined by  $\gamma^{\mu}=e_{a}^{\mu}\left(x\right)\gamma^{a}$ and satisfy  the anticommutation relation $\{\gamma^{a},\gamma^{b}\}=2\eta^{ab}$, where $\eta^{ab}= \mathrm{diag}\left(+ - - \right)$, is the usual  Minkowski  metric. From the spinor theory in curved spaces, the $\gamma^{\mu}$ matrices  must be defined in terms of  field $e_{a}^{\mu}\left(x\right)$ and the Dirac matrices in the space-time, that is, $\gamma^{\mu}=e_{a}^{\mu}\left(x\right)\gamma^{a}$. Note that the $\gamma^{a}$  matrices satisfy the anticommutation relation $\{\gamma^{a},\gamma^{b}\}=2\eta^{ab}$, where $\eta^{ab}= \mathrm{diag}\left(+ - - \right)$, is the Minkowski  metric.  Withal,  the   vielbein fields  satisfy that:  $g^{\mu\nu}=e_{a}^{\mu}e_{b}^{\nu}\eta^{ab}$. Thus, we use the following form  for  the  fields $e_{\mu}^{a} $ and $ e_{a}^{\mu} $ :
\begin{equation} 
e_{\mu}^{a} = \left( \begin{array}{ccc}
1 & 0 & 0  \\
0 & cos \varphi &  -\alpha \rho sen \varphi \\
0 & sen \varphi & \alpha \rho cos \varphi 
\end{array}\right) , \quad  e_{a}^{\mu} = \left( \begin{array}{ccc}
1 & 0 & 0  \\
0 & cos \varphi & sen \varphi \\
0 & -\frac{sen \varphi}{\alpha \rho} & \frac{cos \varphi}{\alpha \rho}  
\end{array}\right) \label{eq32}.\end{equation}
Now,  we use the   Maurer-Cartan structure equation for differential forms , $de^{a}+\omega^{a}_{b}\wedge e^{b}=0$  (condition of zero torsion), where $e^{a}=e^{a}_{\mu}dx^{\mu}$ and $\omega^{a}_{b}=\omega_{\mu\,\,\,b}^{a} \wedge dx^{\mu}$, we obtain $\gamma^{\mu}=e_{a}^{\mu}\gamma^{a}$ , and $\gamma^{\mu}$.   The metric tensor is defined by (\ref{eq28}), and we write   it in terms of the triad as follows: $g^{\mu\nu}=e_{a}^{\mu}\eta^{ab}e_{b}^{\nu}$. We can  choose a dual 1-forms basis, which  describes  the curved  geometry  of  layer  with the presence of topological defect, by $e^{a}=e^{a}_{\mu}dx^{\mu}$, where ${\omega}_{1}^{2} = - {\omega}_{2}^{1} = -(\alpha -1)d\varphi
$. In this way, the  non- zero component of the  spinorial connection is given by
\begin{equation}
\Gamma_{\varphi} = -\frac{i}{2}(\alpha - 1){\sigma}³.
\label{eq34}\end{equation}
Now we consider  the inclusion of  the Aharonov-Bohm   flux  piercing through the centre of the quantum ring. The  vector potential  is given by
\begin{equation}
\mathbf{A} = \frac{\Phi}{2\pi \rho}\widehat{e}_{\varphi},
\end{equation}  
 and adding   the contribution introduced by minimal coupling,   one can write the Eq. (\ref{eq35})  as
\begin{eqnarray} \label{DiracB2}
\gamma^{\mu} \left( i\partial_{\mu} - i\Gamma_{\mu} - e A_{\mu} - \frac{a_{\mu}}{\rho} \right) \psi=0.
\end{eqnarray}
Now, we introduce the confinement potential (\ref{eq1}) in Dirac equation using a coupling similar to Dirac oscillator, thus, the  equation becomes
\begin{eqnarray}\label{DiracB3}
\left[ i\gamma^{t} \partial_{t}
+ i\gamma^{\rho}\left(\partial_{\rho} +\left[ \frac{\sqrt{2a_1}}{\rho} + \sqrt{2a_2}\rho  \right] + \frac{\left(\alpha-1 \right)}{2\alpha\rho}\right)
+\gamma^{\phi} \left( i\frac{\partial_{\varphi}}{\alpha\rho}
+ \frac{1}{\rho}\frac{ e\Phi}{\Phi_{0}} 
-\frac{a_{\varphi}}{\rho}\right)\right] \psi=0,\nonumber \\ 
\end{eqnarray}
where $\Phi_{0}=\frac{|q|}{2\pi}$.  Now , we adopt  the same procedure employed by  Villalba in Ref. \cite{Villa}.  We apply  a  transformation  $S(\varphi)= \exp \left[ -\frac{i\varphi}{2}\sigma^{3} \right] $ to change the representation of Dirac matrices  where the matrices $\gamma^{\rho}$ and $\gamma^{\varphi}$   under similarity transformation $S^{-1}(\varphi)\gamma^{\rho}S(\varphi)=\gamma^{1}  $ and $S^{-1}(\varphi)\gamma^{\varphi}S(\varphi)=\gamma^{2} $ are changed to $\gamma^{1}$ and $\gamma^{2}$,  while we have applied the following transformation in the spin $\psi \longrightarrow S(\varphi) \psi '$.  This transformation  changes the equation (\ref{DiracB3}) to  a new  form  where we  eliminated the  spinorial connection term of the Dirac equation.  Now, we utilize the following  {\it ansatz} for  solution of Dirac equation: 
\begin{eqnarray}\Psi' = e^{-iE t -i\frac{\varphi}{2}{\sigma}³}\left(\begin{array}{c}\psi \\ \phi \end{array}\right) \label{eq44}
\end{eqnarray} 
where $E$ is a  constant. From the similarity transformations and the {\it ansatz} (\ref{eq44}), we obtain the following set of coupled  equations: 
\begin{eqnarray}
E\psi = & -i{\sigma}¹\left[ \frac{\partial }{\partial \rho} + \frac{1}{2\rho} +\frac{\sqrt{2a_1}}{\rho} + \sqrt{2a_2}\rho\right]\phi  -i{\sigma}²\left[  \frac{1}{\alpha \rho} \frac{\partial}{\partial \varphi} + \frac{{ia}_{\varphi}}{\rho} -\frac{i}{\rho}\frac{\Phi}{{\Phi}_0} \right]\phi 
\label{eq47}\end{eqnarray}
and 
\begin{eqnarray}
E\phi = & -i{\sigma}¹\left[ \frac{\partial }{\partial \rho} + \frac{1}{2\rho} -\frac{\sqrt{2a_1}}{\rho} - \sqrt{2a_2}\rho\right]\psi -i{\sigma}²\left[  \frac{1}{\alpha \rho} \frac{\partial}{\partial \varphi} + \frac{{ia}_{\varphi}}{\rho} -\frac{i}{\rho}\frac{\Phi}{{\Phi}_0} \right]\psi 
\label{eq48}.\end{eqnarray}
 By eliminating $\phi$  with use of (\ref{eq48}),  from  (\ref{eq47}) we obtain the following   equation
 \begin{eqnarray}
E² \psi  =
&-&\frac{{\partial}^2 \psi}{\partial {\rho}²} -\frac{ 1}{\rho}\frac{\partial \psi }{\partial \rho} -
\nonumber\\
&-&\frac{1}{{\rho}²} 
\left[\frac{1}{{\alpha}²}\frac{{\partial}²}{\partial {\varphi}²}- \frac{2}{\alpha}\left(\frac{\Phi}{{\Phi}_0} - {a}_{\varphi}\right)\frac{i\partial }{\partial \varphi}-\frac{1}{4}+\sqrt{2a_1} - 2a_1  - \left(\frac{{\Phi}}{{{\Phi}_0}}-{{a}_{\varphi}}\right)^2 \right]\psi +\label{eq49}\\
&+&{\sigma}³  \left[\left(\frac{1}{{\rho}²}-\frac{2\sqrt{2a_1}}{{\rho}²}-2\sqrt{2a_2}\right)\left(\frac{\Phi}{{\Phi}_0}-{a}_{\varphi}+\frac{1}{\alpha}\frac{i\partial }{\partial \varphi}\right)\right]\psi+ 2a_2 {\rho}² \psi +\left( \sqrt{2} +4\sqrt{a_1} \right) \sqrt{a_2}\psi.\nonumber
\end{eqnarray}  
To solve Eq.(\ref{eq49}), we use  the {\it ansatz}
$$
\psi = e^{ij\varphi}\left(\begin{array}{c}R_+ (\rho) \\ R_{-} (\rho)  \end{array}\right),$$ 
where $j= l + \frac{1}{2} $, with $l= 0,\pm 1,\pm 2,..$.  Substituting  $\psi$ in (\ref{eq49}), 
we obtain a  radial equation  for $R_{s}$
\begin{equation}
\left[ \frac{d²}{d{\rho}²} + \frac{1}{\rho}\frac{d}{d\rho} - \frac{{{\delta}_s}²}{{\alpha}²{\rho}²} - 2a_2{{\rho}²}+{\epsilon}_s \right]R_s (\rho) = 0\label{eq50},\end{equation} 
with parameters  ${\delta}_s$, ${\kappa}_s$ and ${\epsilon}_s$ defined as:
\begin{eqnarray}
{\delta}_s &=& {\kappa}_s + \alpha s\sqrt{2a_1}, \nonumber \\ {\kappa}_s & =& \left(l + \frac{1}{2} \right)-\alpha \frac{\Phi}{{\Phi}_0} + \alpha{a}_{\varphi} - \alpha\frac{s}{2},\nonumber \\
{\epsilon}_s & = & {E}² - \frac{2s\sqrt{2a_2}}{\alpha}{\kappa}_{s} - 2\sqrt{2a_2} - 4\sqrt{a_1 a_2} .
\label{eq51}\end{eqnarray}
To solve the equation (\ref{eq50}),  we use  the  change of variable $\xi = \sqrt{2a_2}{\rho}² $, and  obtain
\begin{equation}
\left[ \xi \frac{d²}{d{\xi}²} + \frac{d}{d \xi} - \frac{{{\delta}_s}²}{4 {\alpha}² \xi} - \frac{{\xi}}{4} +\frac{{\epsilon}_s}{4\sqrt{2a_2}} \right]R_s (\xi) = 0 \label{eq53}\end{equation}  
Doing the asymptotic analysis of  Eq. (\ref{eq53}), for the limits  $R_s \rightarrow 0$ and  $\rho \rightarrow \infty$, it is possible
to present radial equation in the following form:
\begin{equation}
R_s(\xi)= e^{\frac{-\xi}{2}}{\xi}^{\frac{|{\delta}_s|}{2\alpha}}F_s (\xi).
\label{eq54}\end{equation}

Substituting the Eq. (\ref{eq54}) in  Eq. (\ref{eq53}), we obtain the following equation for $F_{s}$:
\begin{equation}
 \xi \frac{d²F_s}{d{\xi}²} + \left[ \frac{|\delta |}{\alpha} + 1 - \xi \right] \frac{dF_s}{d \xi}  + \left[\frac{{\epsilon}_s}{4\sqrt{2a_2}} -\frac{|{\delta}_s|}{2\alpha} - \frac{1}{2} \right]F_s (\xi) = 0 \label{eq55},\end{equation} 
 where Eq. (\ref{eq55}) is the hypergeometric equation whose solution is the hypergeometric function:
   \begin{equation}F_s ={\phantom{1}_1}F_1(a,b;z)=
 {\phantom{1}_1}F_1\left(\frac{|{\delta}_s|}{2\alpha} + \frac{1}{2} -\frac{{\epsilon}_s}{4\sqrt{2a_2}},\frac{|{\delta}_s|}{\alpha}+1,\xi\right)\label{eq56}. \end{equation}
 The quantization aspect of the radial solution comes from the fact that hypergeometric functions
must obey a convergence requirement, which is achieved when the first parameter of Eq. (\ref{eq56}) satisfies the condition
\begin{equation}
 \frac{|{\delta}_s|}{2\alpha} + \frac{1}{2} -\frac{{\epsilon}_s}{4\sqrt{2a_2}} = -n.
 \label{eq57}\end{equation} 
Substituting the parameters ( \ref{eq51}) in Eq. (\ref{eq57}), we obtain the energy spectrum for the quasiparticle confined in a quantum ring  in  graphene layer with disclination given by
\begin{eqnarray}
{E}_{n,l}  & = &\pm  \left\{ 4\sqrt{2a_2}\left[n + \frac{|\left(l + \frac{1}{2} \right) - \alpha \frac{\Phi}{{\Phi}_0} + \alpha{a}_{\varphi} - \alpha\frac{s}{2} +\alpha s\sqrt{2a_1}|}{2\alpha} + \right.
\right.\nonumber\\ &+&\left. \left. s\frac{\left(\left(l +\frac{1}{2} \right) - \alpha \frac{\Phi}{{\Phi}_0} + \alpha{a}_{\varphi} - \alpha\frac{s}{2} \right) }{2\alpha} +1 \right]
+
4\sqrt{a_1 a_2}  \right\}^{1/2},
\label{eq58}\end{eqnarray}
where $s = \pm1$ is related with  sublattices $ A/B$ and the eigenvalues  $E>0$ represents electrons  and $E<0$ holes.
Note that in the spectrum, there is the dependence  on  the quantum numbers $n$ and $l$, the parameters $a_1$ and $a_2 $ and the contribution of the non-Abelian gauge field due  to  the presence of the disclination. Note that, in the limit  $\alpha \longrightarrow 1$ we  recover the results of the previous section for a quantum ring in graphene layer. If we consider the quantum dot limit $a_1 \longrightarrow 0$,  we obtain the results obtained in Ref. \cite{Bueno} for  a quantum dot in the presence of topological defect.   In the limit  $a_2 \longrightarrow 0$ we obtain the free quasiparticle case due nature  antidot in the presence of disclination in graphene layer.

 In the previous section, the solutions  with positive energy $E>0$ describe the dynamics of electrons in  the conduction band while  $E<0$  describes the dynamics of holes in the valence band. For  the equation (\ref{eq44}), the spinors corresponding to positive energy are given  by
\begin{eqnarray}
{\Psi}_+ =& f_+ {\phantom{1}_1}F_1 \left(-n,\frac{|{\delta}_+|}{\alpha}+1, \sqrt{2a_2}{\rho}²\right) \times 
\left( \begin{array}{c} 1 \\ 0 \\ 0 \\ \frac{i}{E}\left[2\sqrt{2a_2} \rho + \frac{ \left( {\kappa}_+ - |{\delta}_+| + \alpha \sqrt{2a_1} \right) }{\alpha \rho}\right]
\end{array} \right)\nonumber \\&+ \frac{i f_+}{E}\left(\frac{2n\sqrt{2a_2}\rho}{\frac{|{\delta}_+|}{\alpha}+1} \right){\phantom{1}_1}F_1 \left(-n + 1 ,\frac{|{\delta}_{+}|}{\alpha}+2, \sqrt{2a_2}{\rho}²\right)\left( \begin{array}{c} 0 \\ 0 \\ 0 \\ 1\end{array} \right),
\label{eq60}\end{eqnarray} 
for $s= +1$ and ${\psi}_{-}(\rho) = 0 $, \textbf and by analogy again,
\begin{eqnarray}
{\Psi}_{-} = & f_{-} {\phantom{1}_1}F_1 \left(-n,\frac{|{\delta}_{-}|}{\alpha}+1, \sqrt{2a_2}{\rho}²\right) \times 
\left( \begin{array}{c} 0 \\ 1 \\ \frac{i}{E}\left[2\sqrt{2a_2} \rho - \frac{ \left(  {\kappa}_{-} + |{\delta}_{-}| - \alpha \sqrt{2a_1} \right) }{\alpha \rho}\right] \\ 0
\end{array} \right)\nonumber \\& + \frac{i f_{-}}{E}\left(\frac{2n\sqrt{2a_2}\rho}{\frac{|{\delta}_{-}|}{\alpha}+1} \right){\phantom{1}_1}F_1 \left(-n + 1 ,\frac{|{\delta}_{-}|}{\alpha}+2, \sqrt{2a_2}{\rho}²\right)\left( \begin{array}{c} 0 \\ 0 \\ 1 \\ 0 \end{array} \right),
\label{eq61}\end{eqnarray}
 for $s= -1$ and ${\psi}_{+}(\rho) = 0 $. The factor $f_s = N_s e^{-iEt}$  looks like:
\begin{eqnarray}
 {f}_s =\left( \frac{\sqrt{8a_2}\Gamma \left(\frac{|{\delta}_s|}{\alpha}+n+1 \right)}{\Gamma(n+1)\left[\Gamma \left(\frac{|{\delta}_s|}{\alpha}+1\right)\right]²} \right)^{1/2}  \times e^{-iEt}e^{i(l+1/2)\varphi} e^{-\sqrt{\frac{a_2}{2}}{\rho}^{2}} (2a_2 )^{\frac{|{\delta}_s|}{4\alpha}} {\rho}^{\frac{|{\delta}_s|}{\alpha}}.
 \label{eq62}\end{eqnarray} 
Note that for  $\alpha = 1$, the  non-Abelian gauge field  contribution $a_{\varphi}$ goes to zero. In this way, the parameters ${\delta}_s$, ${\kappa}_s$, ${\epsilon}_s$  reduce to the parameters ${\vartheta}_s$, ${\varsigma}_s$, ${\varepsilon}_s$ respectively. In other words, we recover the case of graphene layer without defect.
\subsection{The persistent current in the presence of disclination}
Now, using the  same procedure  as in  the previous section, we obtain the persistent current for  a quantum ring in  a  graphene layer with  a topological defect. 
We use the expression of  eigenvalues (\ref{eq58}) and the Byers-Yang relation~\cite{Byers} given by 
\begin{eqnarray}\label{current2}
I = -\sum_{n,l} \frac{\partial E_{n,l}}{\partial \Phi}
\end{eqnarray}
In this way,  the persistent current for this system is written as:
\begin{eqnarray}
I = \frac{|q|}{2 \pi}\sum_{n,l} \sqrt{2a_2} \left(  \frac{\pm{\delta}_s}{|{\delta}_s|} +1\right) \times \Bigg \{  4\sqrt{2a_2}\left[ n + \frac{|{\delta}_s|}{2\alpha} + s\frac{{\kappa}_s}{2\alpha} +1  \right] + 4\sqrt{a_1 a_2} \Bigg \}^{-1/2}.
\label{eq59}\end{eqnarray}
So, the persistent current is a  function of the control parameters $a_1$ and $a_2$ and the quantum numbers $n$ and $l$. In addition, we also observed the emergence of correction term ${a}_{\varphi}$.
Note that, the  persistent current is a function of the parameter characterizing the curved geometry introduced by  the topological defect. The expression (\ref{eq59}) is  a function of the non-Abelian magnetic flux and the Aharonov-Bohm flux  presented in the parameter ${\kappa}_s$. The current is an  oscillating function of  the Aharonov-Bohm flux $\Phi_{0}$,  where the oscillation  vanishes for the large values of the  flux. These fluxes are responsible for the  arising  of the persistent current. Note that in the limit $a_1 \longrightarrow 0$ in Eq. (\ref{eq59}) we obtain the persistent current for a quantum dot in the presence of  disclination \cite{Bueno}   given by
\begin{eqnarray}
I = \frac{|q|}{2 \pi}\sum_{n,l} \sqrt{2a_2} \left(  \frac{\pm{\delta}_s}{|{\delta}_s|} +1\right) \times \Bigg \{  4\sqrt{2a_2}\left[ n + \frac{|{\delta}_s|}{2\alpha} + s\frac{{\kappa}_s}{2\alpha} +1  \right]  \Bigg \}^{-1/2}.
\label{eq59a}\end{eqnarray}
In the limit $\alpha \longrightarrow 1$  in Eq. (\ref{eq59a}), we  recover the persistent current for  a quantum ring in a graphene layer obtained in the previous section. Note that we have considered the persistent current for electrons $E>0$. For the case of holes we use similar calculation for the case $E <0$.

\section{Conclusions}
In this  Contribution  we have investigated a quasiparticle in  a quantum ring  in graphene in  two cases with/without the presence of  a topological defect.  We have described  these quasiparticles at low energies by a two-dimensional massless Dirac equation  and have used an exactly solvable theoretical model to describe the  harmonic confinement  in a quantum ring in a graphene layer.  We  claim that our model can  be used  as a theoretical  description of   a quantum ring in  suspended  graphene,   and  in  a future work we can introduce a contribution of  gapped graphene in our model and  study the influence of the presence of gap in this model. We  discussed the dynamics of Dirac spinors  with/without  the impacts of topological defects in the presence of a ring harmonic confining potential. The Dirac oscillator coupling is used with a model for the  ring  with harmonic  potential to confine the quasiparticles.  We have obtained the spectrum of energy of  a quantum ring in the presence of Aharonov-Bohm fluxes  and  find that  the persistent current  is a periodic function  of  Aharonov-Bohm fluxes. The influence of topological defect changes the electronic structure at low energy due to the coupling of the angular momentum with the non-Abelian gauge fields. The energy spectrum depends on the parameter $\alpha$  characterizing the change of the geometry of graphene layer introduced by disclination, and the parameters $a_{1}$ and $a_{2}$   characterizing the confinement potential.  We have obtained the expression for the persistent current  depending on  the parameters characterizing the quantum ring and the topological defect. The current is a periodic function of the Aharonov-Bohm flux and  the non-Abelian flux for case in the presence of  the topological defect. We conclude that the  impact of confinement potential and  geometry introduced by topological defects for the electronic properties of graphene is  crucial for the realization of the future investigation in quantum computation  in these  systems \cite{Knut05,knut5,knut8}.

{\bf Acknowledgements} 

We are grateful to Knut Bakke for interesting discussion. We thank  CNPq, CAPES and CNPq/Universal,  for financial support.


\begin{thebibliography}{99}
\bibitem{novo} K. S. Novoselov, A. K. Geim, S. V. Morozov, D. Jiang, M. I. Katsnelson, I. V. Grigorieva, S. V. Dubonos, and A. A. Firsov, Nature {\bf 438}, 197 (2005).
\bibitem{kim}  Y. Zhang, Y.-W. Tan, H. L. Stormer, and P. Kim, Nature {\bf 438}, 201 (2005).
\bibitem{Novoselov3}M. I. Katsnelson, K. S. Novoselov, A. K. Geim, Nature Physics {\bf 2}, 620 (2006).
\bibitem{Young} A. F. Young, P. Kim, Nature Physics, {\bf 5}, 222 (2009).
\bibitem{stan} N. Stander, B. Huard, and D. Goldhaber-Gordon,Phys. Rev. Lett. {\bf 102}, 026807 (2009); arXiv: 0806.2319.
\bibitem{bula}B. Trauzettel, D. V. Bulaev, D. Loss, and G. Burkard, Nature Physics {\bf 3}, 192, (2007).
\bibitem{Sub}D. Subramaniam, F. Libisch, Y. Li {\it  et al}., Phys. Rev. Lett. {\bf 108}, 046801, (2012).
\bibitem{pono} L. A. Ponomarenko, F. Schedin, M. I. Katsnelson, R. Yang, E. W. Hill, K. S. Novoselov, and A. K. Geim, Science {\bf 320}, 356 (2008).
\bibitem{schn1}S. Schnez, F. Molitor, C. Stampfer, J. G\"{u}ttinger, I. Shorubalko, T. Ihn, and K. Ensslin, Appl. Phys. Lett. {\bf 94}, 012107 (2009); arXiv: 0807.2710.
\bibitem{silv}P. G. Silvestrov and K. B. Efetov, Phys. Rev. Lett. {\bf 98}, 016802 (2007).
\bibitem{schn2}S. Schnez, K. Ensslin, M. Sigrist, and T. Ihn, Phys. Rev. B {\bf 78}, 195427 (2008).
\bibitem{peters1}M D Petrovi\'c, F.  M.  Peeters, A.  Chaves.  and G.  A . Farias, J. Phys.: Condens. Matter {\bf  25}  495301 (2013).
\bibitem{peters2}M. Gruji\'c, M. Tadi\'c  and F. M. Peeters, Phys Rev. B {\bf 87 } 085434 (2013).
\bibitem{peters3}D. R. da Costa, A. Chaves, M. Zarenia, J. M. Pereira Jr.  G. A. Farias, and F. M. Peeters,  Phys Rev. B {\bf 89 } 075418 (2014).
\bibitem{Bleszyski} A. C. Bleszynski-Jayich et al., Science {\bf 326}, 272 (2009).
\bibitem{TanInkson} W.-C. Tan and J. C. Inkson, Phys. Rev. B {\bf 60}, 5626 (1999).
\bibitem{Guinea}B. Wunsch, T. Stauber, F. Sols, and F. Guinea, Phys. Rev. Lett. {\bf 101}, 036803 (2008). 
\bibitem{Ma} M. M. Ma, J. W. Ding and N. Xu, Nanoscale, {\bf 1}, 387 (2009).
\bibitem{eplfur}C. Furtado, A. Rosas and S. Azevedo, EPL {\bf 79  } 57001 (2001).
\bibitem{linco1} L. Dantas and C. Furtado, Phys. Lett. A {\bf 377} 2926 (2013).
\bibitem{linco2}L. Dantas, C. Furtado and  A. L. Silva Netto,  Phys. Lett. A {\bf 379} 11 (2015).
\bibitem{Eckern} U. Eckern and P. Schwab, Adv. Phys. {\bf 44}, 387 (1995). 
\bibitem{Moshinsky} M. Moshinsky and A. Szczepaniak, J. Phys. A: Math. Gen. {\bf 22}, 1817 (1989).
\bibitem{benitez}J. Ben\'{\i}tez, R. P. Mart\'{\i}nez y Romero, H. N. N\' u\~nez-Y\' epez, and A. L. Salas-Brito, Phys. Rev. Lett. {\bf  64}, 1643 (1990).
\bibitem{rozme} P. Rozmej and R. Arvieu, J. Phys. A{\bf 32}, 5367 (1999).
\bibitem{josevi}J. Carvalho, C. Furtado, and F. Moraes, Phys. Rev. A  {\bf 84}, 032109 (2011).
\bibitem{bermu}  A. Bermudez, M. A. Martin-Delgado and A. Luis, Phys. Rev. A {\bf 77}, 033832 (2008).
\bibitem{strange} C. Quimbay, P. Strange, arXiv:1311.2021.
\bibitem{jellal}   A. Belouad, A. Jellal and  Y.  Zahidi, Phys. Lett. A {\bf 380}, 773 (2016).
\bibitem{jellal1}  A. Jellal, A. D. Alhaidari and H. Bahlouli, Phys. Rev. A {\bf 80}, 012109 (2009).
\bibitem{jellal2} H. Bahlouli, A. Jellal and Y. Zahidi, Int. J. Geom. Meth. Mod. Phys. {\bf 11}, 1450036 (2014).
\bibitem{boumeplp} A.  Boumali and H. Hassanabadi, Eur. Phys. J. Plus {\bf 128}, 124 (2013).
\bibitem{franco}  J. A. Franco-Villafa\~ne, E. Sadurn, S. Barkhofen, U. Kuhl, F. Mortessagne and T. H. Seligman, Phys. Rev. Lett. {\bf 111}, 170405 (2013).
\bibitem{Bueno} M. J.  Bueno, J. Lemos de Mello, C.  Furtado and   A. M. de M.  Carvalho, Eur. Phys. J. P {\bf 129} 201 (2014).
\bibitem{Bueno1} M. J.  Bueno, C.  Furtado and   A. M. de M.  Carvalho, Eur. Phys. J. B {\bf 85} 53 (2012).
\bibitem{Bakke} K. Bakke  and C.  Furtado, Phys. Lett. A {\bf 376}, 1269 (2012).
\bibitem{Fock} P. Schluter,  K. H. Wietschorke, W.  Greiner,  J. Phys. A: Math. Gen. {\bf  16} 1999 (1983).  
\bibitem{Aharonov} Y. Aharonov, and D. Bohm, Phys. Rev. {\bf 115}, 485 (1959). 
\bibitem{Abramo} M. Abramowitz and  I. Stegun, \emph{Handbook of Mathematical Functions with Formulas, Graphs, and Mathematical Tables},  chap. 13, pag. 504, Dover Publications Inc., New York, 1964. 
\bibitem{Byers}N. Byers, C. N. Yang, Phys. Rev. Lett. 7, 46 (1961).
\bibitem{kat} M. O. Katanaev and I. V. Volovich, Ann. Phys. (N.Y.) {\bf 216}, 1 (1992).
\bibitem{Lahiri} J. Lahiri, Y. Lin, P. Bozkurt, I. I. Oleynik,  M. Batzill, Nature Nanotechnology, 5, 326 (2010).
\bibitem{Hyde}  S. Hyde, S. Andersson, Z. Blum, T. Landg, S. Lidin,  B. W. Ninham,
\bibitem{Volterra} R. A. Puntingan, H. H. Soleng, Class. Quant. Grav. {\bf 14}  1129 (1997) .
\bibitem{voz1} J. Gonz\'alez, F. Guinea and M. A. H. Vozmediano, Phys. Rev. Lett. {\bf 69}, 172 (1992).
\bibitem{voz2} J. Gonz\'alez, F. Guinea and M. A. H. Vozmediano, Nucl. Phys. {\bf B406}, 771 (1993).
\bibitem{osi1} V. A. Osipov and E. A. Kochetov. JETP Lett. {\bf 73}, 562 (2001).
\bibitem{osi2} V. A. Osipov and E. A. Kochetov. JETP Lett. {\bf 72},  199 (2000).
\bibitem{osi3} V. A. Osipov, E. A. Kochetov and M. Pudlak, JETP Lett. {\bf 96}, 140 (2003).
\bibitem{crespi1} P. E. Lammert and V. H. Crespi, Phys. Rev. Lett. {\bf 85}, 5190 (2000).
\bibitem{crespi2} P. E. Lammert and V. H. Crespi, Phys. Rev. B {\bf 69}, 035406 (2004).
\bibitem{alex2} C. Furtado, F. Moraes and A. M. de M. Carvalho, Phys. Lett. A {\bf 372}, 5368 (2008).
\bibitem{pachos} J. K. Pachos, Contemporary Physics {\bf 50}, 375 (2009).
\bibitem{Villa} V. M. Villalba, Phys. Rev. A {\bf 49}, 586 (1994).
\bibitem{Knut05} K. Bakke, C. Furtado and S. Sergeenkov, Europhys Lett. {\bf 87}, 30002 (2009).
\bibitem{knut5} K. Bakke and C. Furtado, Quantum Inf. Comput. {\bf 11}, 4444 (2011)
\bibitem{knut8} K. Bakke and C. Furtado, Quantum Inf. Process (2012) DOI: 10.1007/s 11128-012-0358-4.


\end{thebibliography}
\end{document}